# Sensing ion channel in neuron networks with graphene field effect transistors

Farida Veliev[1], Alessandro Cresti[2], Dipankar Kalita[1], Antoine Bourrier[1], Tiphaine Belloir[1], Anne Briançon-Marjollet[3], Mireille Albrieux[4], Stephan Roche[5,6], Vincent Bouchiat[1] and Cécile Delacour[1*]

[1]Institut Néel, CNRS & Université Grenoble Alpes, 38042 Grenoble, France

[2]Univ. Grenoble Alpes, CNRS, Grenoble INP, IMEP-LaHC, F-38000 Grenoble, France

[3]Université Grenoble Alpes, HP2 laboratory, Inserm U1042, 38041 Grenoble, France

[4]Université Grenoble Alpes, Grenoble Institut des Neurosciences, Inserm U1216, 38000 Grenoble, France

[5]Catalan Institute of Nanoscience and Nanotechnology (ICN2), CSIC and The Barcelona Institute of Science and Technology, Campus UAB, Bellaterra, 08193 Barcelona, Spain

[6]ICREA – Institució Catalana de Recerca i Estudis Avançats, 08010 Barcelona, Spain

**ABSTRACT.** Graphene, the atomically-thin honeycomb carbon lattice, is a highly conducting 2D material whose exposed electronic structure offers an ideal platform for chemical and biological sensing. Its biocompatible, flexible and chemically inert nature associated with the lack of dangling bonds, offers novel perspectives for direct interfacing with biological molecules. Combined with its exceptional electronic and optical properties, this promotes graphene as a unique platform for bioelectronics. Among the successful bio-integrations of graphene, the detection of action potentials in numerous electrogenic cells including neurons has paved the road for the high spatio-temporal and wide-field mapping of neuronal activity. Ultimate resolution of sensing ion channel activity can be achieved with neural interfaces, and it was shown that macroscale electrodes can record extracellular current of individual ion channels in model systems, by charging the quantum capacitance of large graphene monolayer ($mm^2$). Here, we show the field effect detection of ion channel activity within neuron networks, cultured during several weeks above graphene transistor arrays. Dependences upon drugs, reference potential gating and device geometry confirm the field effect detection of individual ion channel and suggest a significant contribution of grain boundaries, which provide highly sensitive nanoscale-sized sensing sites. Our theoretical analysis and simulations demonstrate that the ion gating of a single grain boundary in liquid affects the electronic transmission of the whole transistor channel, resulting in significant conductance variations. Monitoring the ion channels activity is of great interest as most of neurodegenerative diseases relied on channelopathies, which rely on ion channel abnormal activity. Thus, such highly sensitive and biocompatible neuro-electronics which open the way to FET detection at the sub-cell precision should be useful for a wide range of fundamental and applied research areas, including brain-on-chip, pharmacology, and *in-vivo* monitoring or diagnosis.

**INTRODUCTION.** The long lasting interfacing of neurons with electronic devices is of primary interest for a variety of applications in fundamental neuroscience and biomedical engineering. In particular, the implementation of recording devices to probe the activity of neural circuits at the sub-cellular scale is a critical step towards the understanding of microscopic mechanisms that support information processing. Because ion channels are involved in many neuronal processes or diseases, such as plasticity[1] and channelopathies,[2,3] there is an increasing interest in new technologies able to sense nanoscale events with highly scalable integrated devices on-chip.

Graphene is a promising material for biosciences[4], including interfacing solid-state devices with living cells.[5,6,9,19-21] Using it as the bioelectronics sensing interface can lead to critical improvement on mature solid state technologies, which involve, for example, silicon, transparent ITO or metallic materials. Specifically, the high mobility of charge carriers and the relative chemical inertness to ionic fluids, together with the existence of a 2D electron gas directly exposed on the graphene surface[7] provide unique features when gathered in a single electrode material, which should significantly enhance the detection of extracellular signal when interfacing electrogenic cells on-chip.

Conventional semiconductors such as silicon nanowires require thick insulating layers to reach the best operating regime and to protect the sensors. On the contrary, because graphene is chemically inert, the additional insulating top layer is no more required and a high mobility regime can be accessed in graphene with ultrathin electric double-layers (EDLs), which widely exceeds the threshold performance of conventional semiconductor transistors while keeping similarly high integration and high frequency operation.[8]

Regarding its optical high transparency (97% in the whole visible range) and the ability to transfer macroscale single layer graphene onto a wide range of substrates, including transparent materials, graphene appears as a relevant alternative to the ITO technology for combining optical and electrical addressing for both detection and stimulation.[9]

Moreover, the graphene bendability enables the fabrication of flexible and soft interfaces[10] to well match the mechanical compliance of cells, which is a key feature to enhance the bioacceptance and time reliability of brain interfaces.[11] In addition, graphene itself was shown to favor the healing process by promoting the adhesion and regrowth of damaged neurons.[12] This could limit the proliferation of glial cells around the devices and increase the coupling with the targeted neurons for high quality electrophysiological recordings,[13] thus offering promising perspectives for both brain-on-chip investigations and *in-vivo* monitoring, although immunogenicity and toxicity of graphene sheet should be investigated for clinical applications.[14]

Graphene scaffolds, transistors and electrodes were successfully used for neural tissue engineering,[15,16] regenerative medicine[17,18] and to record neural activity, such as local field potentials *in-vivo* and *in* slice[19-20], isolated spikes in cultured neuron networks,[10,21] and individual ion channels in artificial model systems *in-vitro*.[22] These results have paved the road for monitoring the electrical activity of neuron networks, thus achieving the ultimate detection of ion channels activity, which sustains all information processes.

Here, we have fabricated graphene field effect transistor (G-FET) arrays of different size, starting from microscale graphene channels to compare with previously reported data[22], and then reducing the channel size to reach higher transconductance and transistor sensitivity values. We cultured primary hippocampal neurons above the sensor arrays until their complete electrical maturation (19-21 Days *In-Vitro* DIV), and recorded the evolution of sensors upon drugs, device size and potential gating. Numerical simulations further quantify the impact of grain boundaries, which naturally form in our CVD grown graphene, on the field effect detection of ion channels, and confirm their role in the measured transistor characteristics. Our findings suggest a wide range of subsequent studies for further developing sensing technologies based on graphene bioelectronics.

## RESULTS.

**G-FETs characterizations.** Arrays of graphene transistors with channel width and length of W×L = 1000×250 µm², 40×250 µm², 40×50 µm² and 20×10 µm² (figure 1.a) have been fabricated using high-quality CVD grown monolayer graphene (see Suppl. materials and methods) transferred onto several substrates (sapphire, glass coverslips and silicon on insulator SOI). Some transferred graphene pads were directly contacted by Ti/Au leads by standard photolithography. On the other samples, the graphene sheet was first patterned into smaller channels using photoresist masks and oxygen plasma etching. Source-drain contacts were defined by a second optical lithography step followed by metallization and resist lift-off. Finally, a SU8 resist pattern was used to electrically insulate the metallic contacts from the ionic solution. The isolating resist layer covered the metallic leads only, thus leaving almost the entire sample surface exposed to the neural cell media (figure 1b).

The atomic force micrographs show the absence of surface contamination and the overall quality of the graphene sheet after the device fabrication (figure 1c), which is further confirmed with Raman spectrometry analysis (figure 1d). The intensity ratio between the G and 2D bands (resp. 1583 $cm^{-1}$ and 2676 $cm^{-1}$) $I_G/I_{2D}$ = 0.3 and the width of the 2D-band peak (30 $cm^{-1}$) well match the values reported for monolayer graphene.[23] Moreover, the very low intensity of the D-band peak (1300-1383 $cm^{-1}$) indicates the low amount of lattice defects.

Two-point measurements on graphene stripes with length from 50 μm up to 2200 μm show a linear dependence of the device resistance with the probed channel length (figure 1e). The low value of the measured square resistance is geometry-independent and remains close to $R_{\square}$=600 Ω/square, which demonstrates the overall homogeneity of the graphene material at the mesoscale. The electronic mobility obtained from back gated field effect measurements is estimated to be about 6000 $cm^2.V^{-1}.s^{-1}$ (figure 1f), which is in the higher range for CVD graphene FETs obtained with the same fabrication process.[24]

The liquid-gated G-FET sensitivity is extracted from the transfer curve characterized without cell by applying a DC voltage $V_{LG}$ in the cell culture medium with a quasi-reference Pt electrode while monitoring the drain-source current $I_{DS}$ at fixed bias voltage $V_{DS}$ with a shielded multi-probe station. Input bias voltages $V_{DS}$, $V_{LG}$ were low-pass filtered to reduce electronic noise coming from the electrical setup as illustrated within the figure 2a (but without neuron). Figure 2b shows the conductance for different G-FET sizes. The dependence on $V_{LG}$ shows a symmetric ambipolar field effect behavior with a charge neutrality point around $V_D$ = 0.4V. The drain-source-bias-normalized transconductance $g_m = \partial I_{DS}/\partial V_G \times 1/V_{DS}$ strongly increases when reducing the channel length, since it is inversely proportional to the $L/W$ ratio. The highest value (4 mS/V) is obtained for the smallest devices (20×10 $\mu m^2$), and is comparable with the state of the art.[5,10,21]

**Interfacing matured neurons on-chip.** Primary hippocampal neurons were cultured for periods of 19-21 days on the G-FET arrays, previously coated with poly-L-lysine to enhance the cell adhesion and outgrowth (see Suppl. materials and methods). Scanning electron micrographs provide a close view of the dense and intricate neurite network formed above the graphene transistor channel after 3 weeks (figure 2c). Cultures of pure neurons (without glial cells) are obtained by adding AraC (1 μM) to the serum-free culture medium, which stops the growth and proliferation of glial cells that could remain after tissue dissociation and cell seeding. This is crucial to prevent the formation of an interlayer of astrocytes that could be intercalated between the sensors and the neurons as it would shield the electric activity to be sensed by the G-FETs. However, glial cells are important for the electrical maturation of neurons. Thus, neurons are still maintained in glial-conditioned medium to promote their growth and connectivity. Typically, neurons were seeded at a density of $0.5 \cdot 10^5$ cells/cm². Higher densities were also used for tests on the smallest FETs (20×10 μm²). As shown with the post-mortem immuno-fluorescent labeling (figures 2d-e), neurons exhibit a pyramidal shape, many neurites and synapses, as expected for matured hippocampal neurons.[25]

Cell attached voltage-clamp technique was used to assess the spontaneous electrical activity at the soma of cultured hippocampal neurons at DIV 21 (figure 2f). The shape, amplitude and duration of the detected current peaks can be clearly attributed to inward currents through the cell (typ. $Na^+$ or $Ca2^+$). Additionally, calcium imaging was performed to assess the ionic activities along neurites. Figure 2g shows the typical recordings of calcium fluxes. The signal propagates from the dendrite to the soma and exhibits the expected shape and duration. Thus, both patch-clamp and calcium imaging confirm the cells maturation and operating electrical signaling pathways.

**Electrophysiological recording with G-FETs.** Figure 2a illustrates the typical representation of the neuron gated G-FET. Ionic currents through the cell membrane can be modeled with a parallel membrane capacitance and resistance $C_M$ , $R_M$ and the conductance $\gamma_i$ of the voltage gated ion channels modulated by the membrane potential $V_M$. The opening and closing of membrane channels generate ion currents, which charge the capacitance of graphene and modulate the charge carrier density that is detected by monitoring the drain-source current within the graphene channel.

During the culture, the drain-source current at fixed bias voltage is monitored within the shielded probe station in controlled atmosphere (5% of $CO_2$). The input bias voltages $V_{DS}$, $V_{LG}$ set the G-FETs at maximum transconductance in the hole conduction regime, according to the previous electrical characterizations without cell performed just before the culture. We assess the presence of healthy neurons above the G-FETs with reflective microscope for the opaque silica substrates before the recordings. On the transparent sapphire or glass samples, neurons are observed with conventional transmission microscope.

During the time recordings of neurons-functionalized G-FETs (typically after 19-21 days in culture), the drain-source current of the graphene stripe (which acts as a FET channel) is measured at fixed $V_{DS}$ and fixed $V_{LG}$ and expressed in conductance (figure 3). It shows step-like modulations with well-defined values. The conductance essentially switches between a few (mostly two) constant current values as further proven with the corresponding time-averaged histograms. The shape and time duration of the detected signals do not correspond to the expected signals induced by single spike or local field potential, but are rather well characteristics of the detection of inward and backward ion currents through membrane channels that rapidly switch between two conducting (open/close) states and remain opened or closed during 1 to several milliseconds.

The telegraphic signal induced by neurons is much larger (3 order of magnitude) than the intrinsic electronics noise of the transistor,[10] being around $10^{-16}$ $A^2/Hz$ instead of $10^{-19}$ $A^2/Hz$ without neurons (figure 3d). The Lorentzian bulge superimposed on a $1/f^2$ spectrum in the low frequency regime is characteristic of the noise fluctuations of the ionic current in membrane channels with a mean firing frequency around 30 Hz for neurons.[26]

The signal is completely suppressed if we prevent mechanical motion of the cells by fixing them in 4% paraformaldehyde (figure S1), and also if cell can move but only the ion channel activity is blocked. This clearly demonstrates the strong link with the electrical activity of neurons and excludes an effect of friction or mechanical fluctuation induced by the cells or by the surrounding liquid.

As shown in figure 4, the addition of a prominent sodium channel-blocker (tetrodotoxin, TTX 0.5 μM) to the extracellular medium completely suppresses the signal, thus revealing its unambiguous dependence upon the ion channels activity. Moreover, the incubation with bicuculline (BIC, 20 μM, 15 min 37°C) – a $GABA_A$ receptor antagonist that activates electrical activity – increases the occurrence of the conductance modulations with a switching time varying from few to several hundred milliseconds. Replacing the BIC supplemented medium by fresh culture medium decreases the switching rate, while the subsequent addition of BIC (during the recording at room temperature) partly reestablishes the conductance steps during the time of recording. The reduced activity after the second injection of BIC could be explained by the several medium changes and the lack of the incubation at 37°C, in comparison to the first experiment. Nonetheless, it leads to the emergence of a second excited state (inset of the green trace). The scheme (figure 4b) illustrates the main neurons ion channels and the effect of adding BIC and TTX on their opening and closing dynamics.

The dependence upon the potential gating is illustrated with the figure 5. The polarity of the current modulations inverses from positive to negative conductance peaks once the G-FET is tuned from hole to electron conduction regime, which is in agreement with the expected sign inversion due to the change of the polarity of the charge carriers in the transistor channel. The amplitude of the signal (few μS) remains roughly constant on a wide range of gate potential from $V_{LG}$ = 0V to 0.8V (figure 5b and figure S2), being the highest at the maximum transconductance values and minimum around the Dirac point as expected for a field effect detection (figure 5c). Also, the relative conductance modulation $\Delta G/G$ increases when reducing the device size (figure 6a), as the transconductance becomes higher for smaller G-FETs (figure 2b).

While these features are typical of the field effect detection of individual ion channels, the amplitude of the signals is slightly higher than expected, around 10 to 100 nA in term of drain-source current for larger and smaller devices respectively. Also, another unexpected and clear dependence is observed when reducing the device size. While the sensitivity increases by shortening the transistor channel, the detection efficiency decreases and reaches a threshold for the devices smaller than $W \times L$ = 40×50 μm² (figure 6b). The two-state signals induced by neuron ion channels were never observed on the shortest G-FETs with channel dimensions of $W \times L$ = 20×10 μm², independently of the neuron density. Importantly, increasing the seeding densities to values as large as $1.5 \cdot 10^5$ cells/cm² on more than the 40 samples, did not lead to any appearance of the step-like signals routinely seen for larger FET areas. The absence of signal on those smallest devices does not result from a depression of their performance, since indeed they show a higher transconductance.

## DISCUSSIONS.

A distinctive feature of CVD grown graphene is the presence of highly reactive grain boundaries, which generate a patchwork across the total width of the G-FET channels for CVD graphene[27,28] and provide highly sensitive nanoscale sensing sites.[29] While many efforts were primarily devoted to production of large-scale defect free graphene to reach higher mobility, grain boundaries have appeared as highly sensitive 1D-line defects for analytic applications,[30] allowing for example the detection of individual gasses or adsorbed molecules.[31,32] In particular, Yasaei et al[31] have shown that the sensitivity of a polycrystalline graphene ribbon is 4 times higher than for a pristine (single crystalline) ribbon, and it reaches its maximum value when one single grain boundary crosses the entire width of the G-FET channel. The amplitude of the modulation and its dependence upon the number of grain boundaries, convincingly match our observations (figures 6a and 6b). The modulation of the conductance $\Delta G/G$ strongly increases when reducing the device size, reaches its maximum value when a single grain boundary crosses the transistor width, and then falls sharply when the G-FET channel becomes shorter than the grain size. Since the average grain size is about 30μm within our graphene, no grain boundary is expected to cross the 10-μm-long graphene FET channels. The consequence could be the strong suppression of the conductance modulation.

The size of a single grain was estimated by stopping the CVD growth shortly before the complete coalescence of single grains. This method gives an accurate and large scale evaluation of the nucleation density and size of the mono-crystalline grain being around 30 μm (figures 6c-d). Because we are using a pulsed growth process, the grains have irregular shape, but their nature remains that of a single crystal. This can be also confirmed by increasing the time of the growth, because multilayer patches mark the nucleation center of the each single grain.

We estimate the contribution of the GBs to the global conductivity by rescaling the G-FET conductance with the device size. For a sample $A$ without GB (20×10 μm), the conductance scales as $G_A = \sigma\, W_A/L_A$, where $\sigma \equiv G_A\, L_A/W_A$ is the single-grain conductivity of the pristine graphene (without GB) and we neglected the contact resistance. For a sample *B* with possibly one enclosed GB (40×50μm), the contribution of the single grains to the global conductivity is $G_{SG} = \sigma W_B/L_B = G_A L_A/W_A W_B/L_B$, in this case $G_{SG}$=2/5 $G_A$. After aligning all Dirac points of the different curves at the same energy, we select three different $V_G$ values and calculate $G_{SG}$. $G_{SG1}$=0.3mS × 2/5 = 0.12 mS, $G_{SG2}$ = 0.32 mS and $G_{SG3}$ = 0.40 mS (figure 6e). The discrepancy with the experimental value (0.09 , 0.25, 0.39 mS), determines the contribution of a single grain boundary, being the highest (25%) at lowest energy. Such contribution is relatively small compared to that of the other sources of disorder, as also observed by Huang et al.[28]

**Impact of grain boundaries.** To strengthen the hypothesis of the large contribution of grain boundaries on the field effect and its impact on the detection of single ion channels in liquid, we performed a theoretical analysis based on the following experimental facts: (i) based on its time scale and its suppression in the presence of a chemical inhibitor, the signal is generated by the neuron activity; (ii) the signal is visible *only* in samples where grain boundaries are present; (iii) the signal roughly varies between two states, with a conductance decrease on the hole-side of the spectrum when the ion channels open, and opposite behavior on the electron side; (iv) the signal is generally stronger on the hole-side of the spectrum and can be absent or degraded depending on specific samples.

A possible explanation for the observed behavior is that the grain boundary acts as a potential barrier whose height is modulated and inversed by the ion charges ejected from the cell channels (figure 6f). This requires the potential barrier to be high and large enough to significantly affect the global electron transport, and that the ions to rapidly diffuse along the grain boundary, to allow that the resistance variation follows the time scale of the neuronal activity. This assumption is supported by a recent simulation study,[33] based on density functional theory, which has demonstrated that the diffusivity of $Na^+$ ions along grain boundaries is about one order of magnitude larger than inside a graphene crystal. We thus hypothesize that grain boundaries effectively behave as ion traps for motion of the ions, which are successively reabsorbed by the cell once the active pumps start working.

The high reactivity of the grain boundaries and the environment of the liquid gate might also enhance the charge accumulation and the resulting potential barrier in the absence of the neurons. Radchenko et al.[34] have modelled grain boundaries as unidimensional charge lines in the graphene plane, and estimated the potential barrier within the self-consistent Thomas-Fermi approximation. Inspired by this work, we propose here a simple theoretical model and numerical simulations to support our experimental data and our interpretation. We consider a large graphene ribbon over a $SiO_2$ substrate and covered by water (corresponding to the liquid gate, which would require a much sophisticated model to be accurately described[35]), see the inset of figure 6g. A metallic back-gate tunes the charge density in the graphene layer. In the experiment, the gating is performed through the ion liquid, but again this is not essential for our purposes and it would introduce a further complication. The grain boundary is described by a linear charge distribution with density $\lambda$. When the ion channels are closed, we assume a negative density $\lambda_c$=-*e*/nm. The exact value of $\lambda_c$ is not critical for a qualitative discussion. When the ion channel opens and the $Na^+$ ions are trapped along the grain boundary, we assume an opposite positive charge density $\lambda_o$ =*e*/nm. We place the charge 0.5 nm above graphene and consider a lateral extension of 2 nm and a height of 0.1 nm. The choice of these parameters is rather arbitrary, and is intended to mimic the effect of charged molecules and ions grafted along the grain boundary, which are expected to be just above graphene and distributed over a certain width. The system is considered as periodic along the GB direction. The electrostatic potential is calculated by self-consistently coupling the 2D Poisson equation and the charge density as obtained from the Thomas-Fermi approximation. We assume a linear density of states for graphene, which is a reasonable approximation around the charge neutrality point. As for the Poisson equation, we consider Dirichlet boundary conditions on the gate and Neumann boundary conditions on the other edges of the simulation box. We proceed by first considering the pristine case ($\lambda$=0) and calculate the electrostatic potential, the charge density and chemical potential (with respect to the charge neutrality point) as a function of the gate potential. Then, we repeat the calculation in presence of the linear charge distribution. The difference between the electrostatic potential obtained in the two cases allows us to extract the potential barrier seen by electrons as a function of the chemical potential.

Figure 6g shows the potential profile calculated for some representative chemical potentials at T=300 K and for the positive linear charge density $\lambda_o$=*e*/nm. The results are analogous but reversed in sign for the negative linear charge density $\lambda_c$=-*e*/nm. We note that the screening is much less effective when the chemical potential is close to the charge neutrality point, because of lower charge density. Another interesting point is that due to the high permittivity of the water, the potential barrier is weaker but more spatially extended than if the line charge were buried in the oxide.

By using the obtained potential profiles, we calculate the differential conductance for a *W*=25 nm wide graphene ribbon with a grain boundary across its width, see the inset of figure 6h. The simulations are based on the non-equilibrium Green's function approach[36] and consider a minimal first-nearest-neighbor tight-binding model to describe graphene.[37] As shown in figure 6h, when the ionic channels are closed, the positive potential barrier strongly suppresses the conductance for electrons close to the charge neutrality point, due to the formation of an n-p-n junction. At larger chemical potentials, the differential conductance is decreased compared to the pristine case, with a general shift induced by the lower density of states in the grain boundary region. For holes, the conductance is not significantly affected, because the potential in the grain boundary region induces a higher density of states than in the rest of the system. When the $Na^+$ ions are trapped along the grain boundary, the situation is reversed, with a lower conductance on the hole side. As a consequence, the opening of the ion channels has the effect to increase the conductance for holes and decrease the conductance for electrons, as observed experimentally and indicated by the arrows. We note that different values of $\lambda_c$ and $\lambda_o$ do not affect our qualitative picture. For example, a smaller charge density $\lambda_c$ would just entail a weaker signal on the electron side, as indeed experimentally observed. Even in the case $\lambda_c$=0, the conductance decrease induced on both sides of the spectrum by the line defects along the grain boundary can be modulated by the $Na^+$ induced barrier (see Suppl. Materials figure S3 and methods). The essential mechanism at the base of our interpretation is the enhanced diffusivity of the ions along the grain boundary.

A remarkable experimental aspect is the roughly constant amplitude, at given gate potential, of the signal modulation. This indicates that the charge along the grain boundary basically has two given values depending on the state of the ion channel, and then that the $Na^+$ ions saturate the traps. This is reasonable if we consider that the total number of ejected ions *per* channel is in the order of $5x10^5$, and that the length of the grain boundaries is limited by the sample width to few tens of μm.

We finally note that simulated conductance variations are larger than those experimentally observed. However, the adopted model does not consider the presence of (the rather strong) bulk disorder, which acts an additional resistance within the monocrystalline grains.

Other explanations of the observed conductance modulation are possible, which, however, we can easily exclude. For example, the observation of step-like conductance modulations in FETs has also been related to the electronic noise intrinsic to the G-FET. Finite number of impurities in close vicinity of the conductive channel,[38] at the interface with the substrate or the top gate, can trap the charge carriers thus resulting in conductance fluctuations. Upon increasing the number of impurities, random telegraph signals superimpose into a single 1/f noise spectrum, as observed without neurons. If only one prevalent impurity were present close to the FET channel, the conductance would fluctuate between two discrete values corresponding to individual trapping/detrapping events. This is usually observed in nanoscale 1d or 2d devices, such as silicon nanowires,[39,40] carbon nanotubes[41] and graphene[42] at low temperature where the thermal fluctuations are almost suppressed and the current is carried by a small number of charge carriers concentrated in a small region. However, this cannot explain our signals induced by neurons, which rather exhibit a high conductance modulation on large graphene transistors (W×L = 1000×250 µm²; 40×250 µm²; and 40×50 µm²), while no modulation was observed for the smallest one (20×10 µm²).

Other defects could contribute for the interaction with the neurons. Gating of edge states have been shown to play a critical role for the charge carriers transport in graphene nanoribbons, resulting from rough edges, imperfections of the graphene layer or the broken symmetry of the hexagonal lattice and change in bonding. Regarding the unperfected lithographical, etching and transfer techniques, edge states may exist in our G-FETs. However this scenario cannot explain our results, such as the detection threshold observed for smaller G-FETs, and the highest detection probability obtained with larger devices with no edge (figure 6b),

Unfortunately, it is not possible to track ion trapping, neither to localize nor characterize the transmission of individual grain boundaries gated by neurons. Similar experiments with model systems could provide new insights into this non-conventional (heterogeneous) field effect detection with grain boundary, by comparing the responses of single- and poly-crystalline G-FETs for instance. Also, because graphene is optically transparent it would be possible to combine G-FETs monitoring with optical stimulations by using optogenetics, and investigate further the dynamics of specific ion channels.

**CONCLUSION.** We have shown that graphene field effect transistors enable the detection of ion channel activity within hippocampal neuron networks cultured in-situ during 3 weeks on the biosensor array. The detection of ion currents is stable in time and sensitive to neurotoxin affecting the activity of the membrane channels. G-FETs turn out to be innovative and promising platforms for monitoring fundamental electrophysiological processes in living cells at the ultimate resolution of single ion channel sensing. The high carrier mobility within graphene enhances the device response, while the device can be reduced to localize the detection within the neurons network, thus confirming the advantage of G-FETs in comparison to previous large electrodes. Also, our observations suggest that grain boundaries could significantly contribute to the detection, both by acting on the electron transmission and enhancing the ion trapping and diffusion onto the transistor channel. This ultimate resolution is of primary interest when interfacing neurons, because all processes, in healthy or diseased networks, rely on the activity of ion channels. Being highly neuron-compatible and transferable on a wide range of supports, graphene-based bioelectronics offers promising perspective for monitoring the dynamics of neuron-based circuits in pharmacology, medicine and fundamental neurosciences.

**Supplementary materials and methods.**

***CVD graphene growth and transfer onto arbitrary substrates.*** High-quality monolayer graphene was grown on copper foil (25μm thick, 99.8% purity, Alfa-Aesar) using pulsed chemical vapor deposition (CVD) as reported earlier. [24] Pulses of $CH_4$ (2 sccm 10s, then 60s off) are injected into the growth chamber with hydrogen atmosphere. Continuous $CH_4$ flow usually results in an increasing amount of carbon atoms dissolved in Cu foil defects. The following segregation of carbon atoms to the surface of the Cu foil leads to an uncontrolled formation of graphene multilayers. In contrast, using pulsed $CH_4$ flow the copper foil is periodically exposed to pure hydrogen, which binds the segregated/dissolved carbon atoms and carries them out from the growth chamber, preventing the development of multilayer patches. Before the growth, Cu foil is cleaned in acetone and annealed in diluted $H_2$ atmosphere (dilution in Ar at 10%) at 1000°C for 2h. Pieces of Cu foil of about 4×4 mm² with graphene layer grown on top are covered with PMMA on the graphene side and then wet etched in ammonium persulfate solution (0.1 g/ml, 2h at room temperature). After complete etching of Cu, graphene-PMMA stack is rinsed in 6 subsequent deionized (DI) water baths to remove the residual etchant. Then the graphene-PMMA film floating on the DI water surface is scooped from below onto a clean substrate and dried at room temperature. Finally PMMA is removed in an overnight acetone bath followed by a 3 h long thermal annealing at 300°C in vacuum.

**Mobility extraction for electrical measurement.** To extract the mobility different scattering mechanisms, such as elastic scattering including charge impurities, neutral defects and charge transfer from doping, substrate or surface roughness, as well as inelastic scattering on graphene or substrate phonons, should be taken into account.[43] However, the fitting technique[44] we used gives a good approximation. One can extract the mobility in the diffusion region using the conductivity G and the charge carrier concentration, $G = 1/(R - R_c) = \sigma W/L = ne\mu W/L$ and $n = \sqrt{n_0^2 + [n_{unit}(V_G - V_D)]^2}$, with $n_{unit} = \varepsilon_{SiO2}.\varepsilon_0/d.e \sim 7.56.10^{10} cm^{-2}.V^{-1}$ for 285nm $SiO_2$. From this, the graphene resistance R is $R = \frac{L/W}{e\mu\sqrt{n_0^2+[n_{unit}(V_G-V_D)]^2}} + R_C$ where W/L is the width divided by length (aspect ratio) of the device, $R_c$ is the contact resistance, and $n_0$ the residual carrier concentration. Using this formula we can fit the mobility μ, $R_c$ and $n_0$ by measuring the resistance-gate curve as shown in Figure 1f.

***Cell culture and immunofluorescence imaging.*** Primary hippocampal neurons were dissociated from E16.5 mouse embryos and seeded with a density of $0.5 \cdot 10^5$ cells/cm² onto sterilized poly-L-lysine coated chip surface following previously reported culturing protocol.[12] A PDMS chamber (200-300μl) is fixed on the chip for containing the cells and mediums, while keeping dried the deported contacts. The seeded neurons were incubated at 37°C and 5% $CO_2$ in the attachment medium (MEM supplemented with fetal bovine serum) and replaced 3 to 4 hours later by glial conditioned Neurobasal medium supplemented with AraC (1 μM) to stop proliferation of glial cells. Medium was changed once a week. Presence of cells above the device is checked before the measurement with a reflection microscope (Olympus BX51) for silica substrate. On the sapphire substrate, the neurons growth is observed all along the culture with conventional transmission microscope providing better resolution of the neurites position above the device. After recordings, immunofluorescence staining was performed to locate the cells above the devices and characterize their maturation stage. Neurons were fixed in 4 % paraformaldehyde (10 min) and immunostained with Phalloidin, DAPI and anti-synapsin primary antibody to visualize the actin filaments, nucleus and pre-synaptic vesicles respectively.

**Patch clamp and calcium signaling in cultured hippocampal neurons.** After 21 days of culture, 2-5 μl of Fluo-4 AM was added to the cell culture medium depending on the neuron density and incubated for 15 - 60 mins. Then the cell culture medium was changed to remove the excess Fluo-4 AM molecules, and after a short incubation, the neurons were excited by $\lambda_{exc}$ = 488 nm laser light, and the emission at the wavelength $\lambda_{em}$ = 515 ± 15 nm was detected. The data were acquired at 4.76Hz using a confocal microscope and commercial software EZ2000 (Nikon).

**Simulation approach.** The simulation of electron transport in a ribbon with a charged grain boundary consists of two phases: first we determine the electrostatic potential and then we proceed with the quantum transport simulation itself.

For the self-consistently determination of the electrostatic potential, we consider the configuration illustrated in the inset of figure 6(g), where the system is periodic along the *y*-axis and the graphene layer is considered as an *xz*-surface at positon $z = z_G$. We solve the 2D Poisson equation for the electrostatics

$$\nabla_{xz}\epsilon(z) \cdot \nabla_{xz}\varphi(x,z) = -\omega(x,z) \,,$$

where $\nabla_{xz}$ is the gradient in the *xz*-plane, $\epsilon(z)$ is the dielectric permittivity, $\varphi(x,z)$ is the electrostatic potential and $\omega(x,z)$ is the charge density, which includes the charge along the grain boundaries and in graphene. We consider Dirichlet boundary conditions on the bottom gate at position $z=z_G$, with φ□ w $z_G$□=$V_G$ , and Neumann boundary conditions on the other surfaces of the simulation domain, that meaning zero *z*-component of the electric field on the top *xy*-surface and zero *x*-component of the electric field on the side *yz*-surfaces. The Poisson equation is self-consistently coupled through the potential φ to the Thomas-Fermi approximation for the surface charge density η in 2D graphene

$$\eta(\mu,T,x) = -e\int_0^{\infty}\rho(E)f(E,\mu+\mathrm{e}\varphi(x,z_G),\tau)dE + e\int_{-\infty}^{0}\rho(E)[1-f(E,\mu+\varphi(x,z_G),\tau)]dE,$$

where μ is the chemical potential, $\tau$ is the temperature, $f$ is the Fermi-Dirac occupation function, and we considered the linear approximation for the density of states in graphene $\rho(E) = 2|E|/(\pi\hbar^2 v_F^2)$, with $v_F$ is the Fermi velocity at the Dirac points. The two terms on the right side of the equation correspond to the electron and hole contributions, respectively.

The simulation strategy consists in first calculating the electric potential for the system without charge along the grain boundary. If we fix $\mu$=0, the potential energy for electrons on graphene $u_0(x) = -e\varphi(x, z_G)$, which indeed is independent of $x$, i.e. $\varphi(x, z_G) = \varphi_0$, and corresponds to the position of the charge neutrality point with respect to the chemical potential. Therefore, if we consider the energy reference at the charge neutrality point of pristine graphene ($E$=0), then the effective chemical potential becomes $\bar{\mu} = -u_0 = e\varphi_0$, which is directly related to the charge density in the system, i.e. the charge vanishes for $\bar{\mu} = 0$. Then, we repeat the calculation in the presence of the charge along the grain boundary and obtain the spatially varying potential $\varphi(x,z_G)$ on graphene. From these results, we extract the energy potential barrier seen by electrons in graphene, which is given by $U(x) = -e\,[\varphi(x, z_G) - \varphi_0]$. Some typical profiles are shown in figure 6(g) and discussed in the main text.

The second phase of the numerical simulation consists in describing the graphene ribbon by a single orbital spin-degenerate tight-binding Hamiltonian of the type

$$H = \sum_i U(x_i)\ |\psi_i><\psi_i| + \sum_{\langle ij\rangle} t\,|\psi_i><\psi_j|\ ,$$

where $|\psi_i>$ is the ket representing the orbital $p_z$ on the carbon atom with index $i$ and the coupling $t$=-2.7 eV is limited to couples of first-neighbor atoms with indices <ij> . The on-site energy $U(x_i)$ corresponds to the previously calculated energy potential barrier in correspondence of the position of the atom $i$. Finally, the Hamiltonian is passed to a home-grown code for the simulation of quantum transport, which is based on the nonequilibrium Green's function technique. This approach considers the system of interest within two semi-infinite periodic leads, which play the role of source and drain contacts. For small source-drain bias, the transmission coefficient $T$ as a function of the electron energy $E$ is given by

$$T(E) = \mathrm{Tr}\,[\,\Gamma_S(E)\, G^R(E)\, \Gamma_D(E)\, G^A(E)]\,,$$

where $\Gamma_S(E)$ and $\Gamma_D(E)$ are the matrices of the rate operators for the source and drain contacts, respectively, $G^R(E)$ and $G^A(E)$ are the retarded and advance Green's function, respectively, and Tr is the trace operator. For further details, we refer the reader to ref. [36]. The differential conductance at finite temperature $\tau$ and at chemical potential $\mu$ is finally obtained as

$$G(\mu,\tau) = \frac{2e^2}{h}\frac{1}{4k_\mathrm{B}\tau}\int dE \quad \frac{T(E)}{\cosh^2[(E-\mu)/(2k_\mathrm{B}\tau)]}\,,$$

where $2e^2/h$ is the conductance quantum, $k_\mathrm{B}$ is the Boltzmann constant, and cosh is the hyperbolic cosine.

**Simulation of transport across grain boundaries with different geometries.** In the literature, grain boundaries are often modeled by taking into account their specific geometry with the first-nearest neighbor tight-binding Hamiltonian description and disregarding the presence of a potential barrier. While this is not fully justified in our case due to the high GB reactivity associated with the ion liquid graphene is immersed in, we explore anyway this possibility, which could operate jointly with the mechanism illustrated in the main text.

We considered six different grain boundary geometries, see figure S3(a), and calculate the transmission coefficient as a function of the energy for a 100 nm wide graphene ribbon where GBs are transversally placed, see figure S3(b). Note that, depending on the GB geometry, the lattice orientation can be different on the two sides of the ribbon. Our tight-binding Hamiltonian, see the previous section, does not take into account the variation of the interatomic distance with respect to the pristine case. This approximation does not affect our conclusions. The transmission coefficient is rather strongly affected by the grain boundaries and significantly varies for different GB geometries. In most of the cases, we observe a strong decrease of the transmission coefficient around the charge neutrality point and a marked electron-hole asymmetry, which is imputable to presence of odd-numbered carbon rings along the GB.

Since the GBs in the experimental samples are much longer that 100 nm, we can expect that different line defect geometries occur appear along their length. In order to qualitatively account for that, figure S3(c) shows the average transmission coefficient over the six different configurations. The result is more regular with a less pronounced electron-hole asymmetry. To test the effect of a local potential variation and mimic the effect of trapped ions, we added a potential barrier $U$=-50 meV and $U$=100 meV along the GB. The results show a non-negligible variation of the transmission coefficient away from the charge neutrality point. The variation is opposite for electrons and holes, and for different sign of the potential barrier. At negative energies close to the charge neutrality point, the transmission coefficient does not significantly vary.

While these results are quite sensitive to the specific GB geometry, they are compatible with the experimental results and their contribution cannot be excluded. Our conclusions are not affected, since, as indicated in the main text, the essential ingredient turns out to be again the chemical reactivity of the GB and the resulting potential barrier, which modifies the conductance by acting as a local gate.

## FIGURES

**FIGURE 1. G-FETs fabrication and characterizations.** Schematics (a) and optical micrographes (b) illustrating the GFETs with channel width and length varying from 1000×250 μm² to 10×20 μm², from left to right. The contact lines are colored in red, and the graphene active area in grey. (c) Atomic force micrograph of the transistor channel. (d) Raman spectrum of the graphene channel with characteristic peaks ($\lambda_{exc}$=532nm). (e) Determination of the square resistance of graphene using 40 μm wide graphene stripes with varying length *L*. (f) Field effect curve measured on a 40×60 μm² G-FET in ambient environment. The conductance is modulated by a p-doped Si back-gate with 285 nm thick $SiO_2$ gate oxide. Data (dots) are plotted as resistance versus back-gate voltage and fitted to extract the carrier mobility.

**FIGURE 2. Interfacing matured neuron with G-FET.** (a) Equivalent electrical circuit of neuron gated G-FET. The transistor is voltage-biased $V_{SD}$ while the drain-source current $I_{SD}$ is recorded. The conductance is then expressed within all the figures. The liquid gate potential $V_{LG}$ is set using a reference electrode RE. $C_{EDL}$ is the capacitance of the EDL formed at the graphene/liquid interface and $R_{cleft}$ is the seal resistance between the cleft and the bath. (b) The left panel shows the transfer curves for G-FETs with different width-to-length ratios of the transistor channel: *W*×*L*=20×10 μm² (green line), 40×50 μm² (red line), 40×250 μm² (blue line) and 40×750 μm² (black line), and with the Dirac point $V_D$=0.406 V. The right panel shows the non-linear dependence of the normalized transconductance of G-FETs with the number of squares (*L/W* ratio). (c) Representative scanning electron micrograph of 21 days old hippocampal neurons interfaced with the G-FETs. The 250 μm-long graphene strip appears darker between the two bright metallic electrodes. Representative optical images of the neurons stained with phalloidin (d) and synapsin (e). Scale bars are 50 μm. (f) Patch clamp recording on neuron soma

(21DIV) with a closed view of inward $Na^+$ current peak. (g) Optical micrograph (left) of 21 DIV neuron loaded with calcium sensitive fluorescent molecules (Fluo-4). The arrow represents the neurite along which the intensity change of the calcium signal was measured (from red to pink traces respectively). Right. Recording of calcium signals - expressed in term of relative change of fluorescence intensity $\Delta F$/F - along the path indicated by the arrow on the left optical micrograph. The detected peaks correspond to an increase in the intracellular $Ca^{2+}$ ion concentration.

**FIGURE 3. In-situ neurons recording with the several G-FET arrays.** Typical time-traces and histograms of the G-FETs conductance interfaced with electrically matured neurons (19DIV). (a) W×L= 1000×250 μm², $V_{SD}$ = 50 mV, $V_{LG}$ = 0.2 V; (b) W×L = 40×250 μm², $V_{SD}$ = 30 mV, $V_{LG}$ = 0.15 V, and (c) W×L = 40×50 μm², $V_{SD}$ = 30 mV, $V_{LG}$ = 0 V (Devices layout detailed in figure 1a). Optical and immuno-fluorescence micrographs show the presence of healthy neurons above the sensors for the samples. Neurons are stained with synapsin (grey/green) and Dapi (blue/red). d) Current noise power spectral density of the telegraphic signal induced by neurons, with $1/f^2$ and 1/f noise dependence (red/green line) for comparison. The arrow indicates the specific Lorentzian kink. Data are taken from the conductance trace shown in (c).

**FIGURE 4. Dependence upon drugs.** Subsequent conductivity time traces and corresponding histograms recorded on the same G-FET (1000×250 μm²) interfaced with neurons. $V_{SD}$ = 50 mV, $V_{LG}$ = 0 V. The culture medium was first incubated with bicuculline (BIC, black trace), then replaced by fresh medium (washed, blue trace), once again supplemented with bicuculline (BIC, green trace) and finally tetrodotoxin (TTX, red trace) was added to the medium. The arrows indicate the emerging of stable two-state conductance behavior once BIC is added. The bottom

insets show the zoomed view at the position of the second peak of the current histogram, which corresponds to the opened state. b) Schematic diagram illustrating the expected impact of BIC and TTX on the activity of ion channels. From left to right : sodium ($Na^+$) and potassium ($K^+$) leak and voltage-gated channels, ligand-gated chloride $Cl^-$ channels, small conductance calcium activated potassium channels (SK) and the ion pump.

**FIGURE 5. Dependence upon the potential gating.** a) Optical micrograph of neurons (21DIV) on the 40x250µm G-FETs. Metallic electrodes appear yellow, phalloïdin and synapsin staining appear red and green respectively. Scale bar is 40 µm. b) Conductance time trace recorded for several gate offsets ($V_{LG}$ = 0.15, 0.3, 0.5, 0.8 V) and (c) extended peaks for the selected gate offsets showing the polarity change while the device is tuned from the hole ($V_{LG}$ < 0.4V) to the electron conduction ($V_{LG}$ > 0.4V) regime. (d) Conductance change for the several gate offsets (counted peaks N = 32, 96, 39, 46 respectively). The peak was selected in means of signal duration (equal or above 1 ms).

**FIGURE 6. Scaling effect on the G-FET response.** (a) Relative change of conductance $\Delta G/G$ and (b) detection efficiency of the telegraphic signal induced by neurons as function of the area of the graphene transistor channel. The observation probability was obtained using several independent cultures and G-FET arrays with identical culturing and sample fabrication protocols. For W×L = 1000×250 µm² - 5 devices out of 15 tested exhibited RTS when interfaced to neurons, for W×L = 250×40 µm² – 2 out of 28 tested devices, for W×L = 50×40 µm² - 2 out of 40 tested devices, and for W×L = 20×10 µm² – 0 out of 40 tested devices. Representative scanning electron (c) and optical (d) micrographs showing the expected GB network (brightness area). The CVD growth (on Cu foil) was stopped shortly before the complete coalescence of single graphene grains (darkest area). Scale bars 15µm. e) Contribution of one grain boundary to the global conductance estimated from the conductance of a G-FET without GB (black line) and with one single GB (red line). Data are taken from figure 2b. f) Illustration of randomly opening and closing ion channel above a graphene grain boundary crossing the transistor channel. Bottom, schematic representation of a p-n-p junction formed across the graphene grain boundary. Ionic currents flowing through the ion channel tune the Fermi level of the grain boundary, resulting in varying transmission properties. (g) Simulated potential barrier in the presence of a line charge with positive density $\lambda$=e/nm at T=300K and for different chemical potentials. Inset: system configuration for the determination of the potential barrier. (h) Simulated differential conductance G and differential conductance per unit of width G/W for a W=25 nm wide ribbon for $\lambda$=±e/nm and $\lambda$=0, at T=300K. The arrows indicate the conductance change when the ion channels open. Inset: sketch of the simulated ribbon.

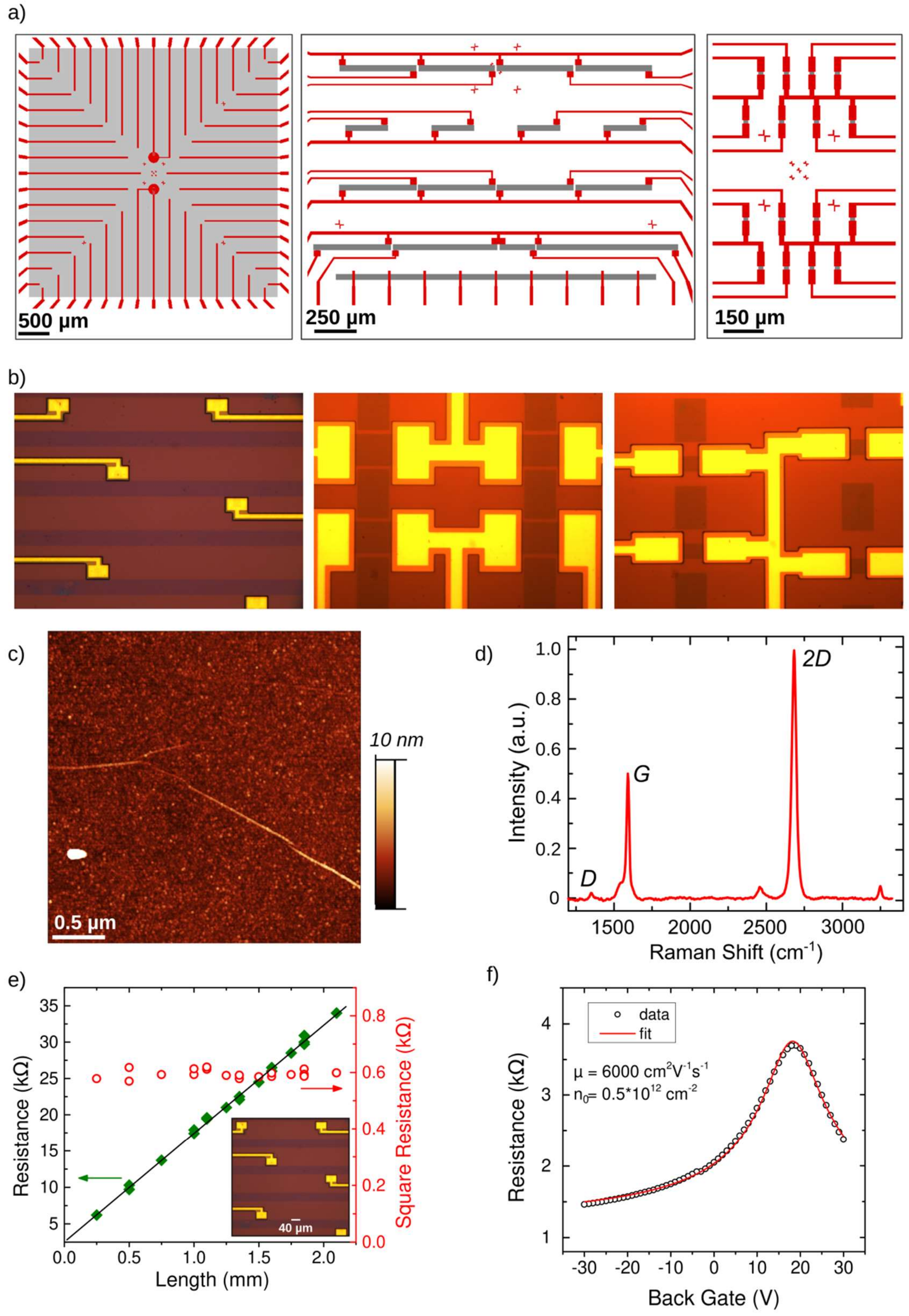

Figure 1.

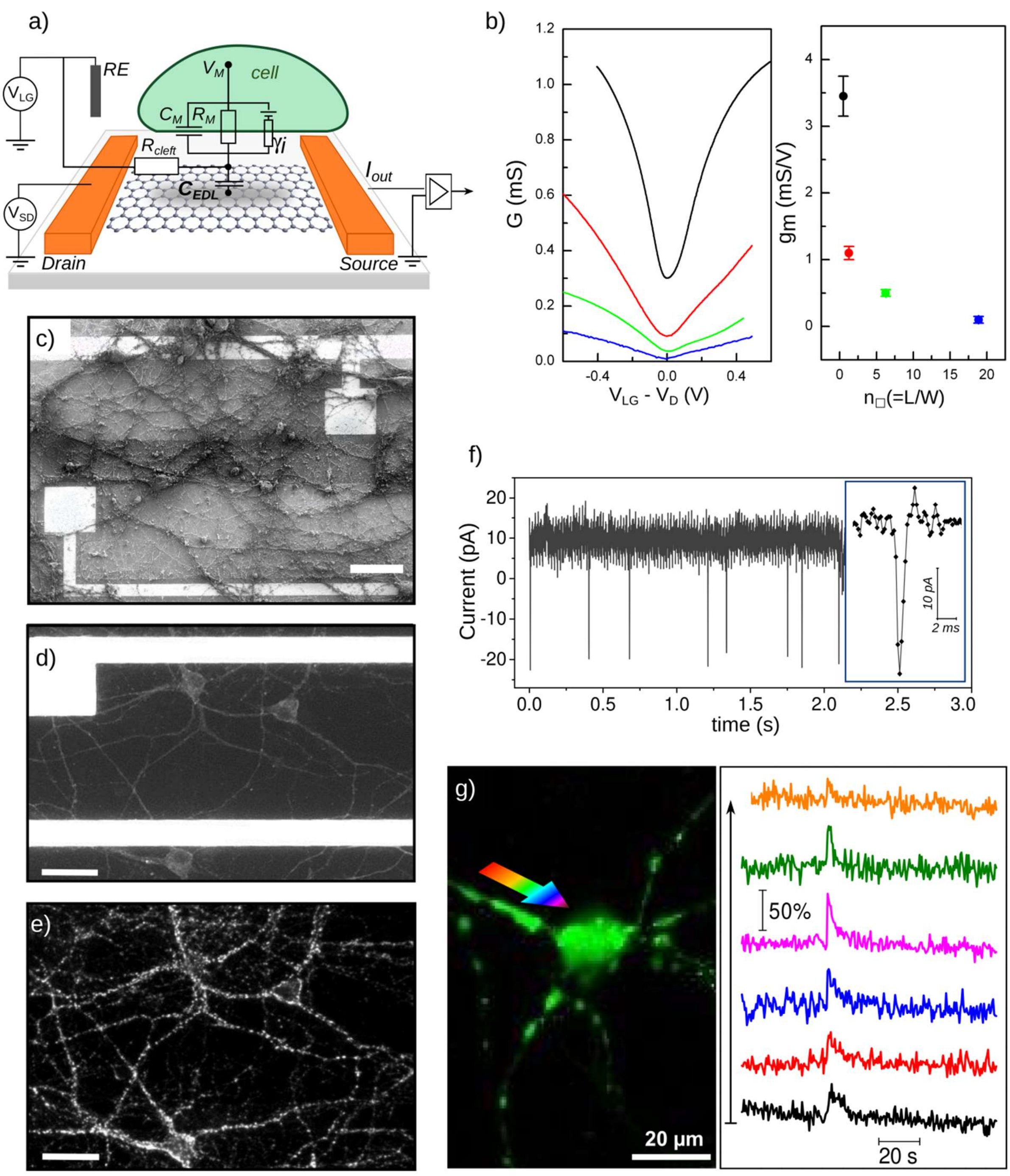

Figure 2.

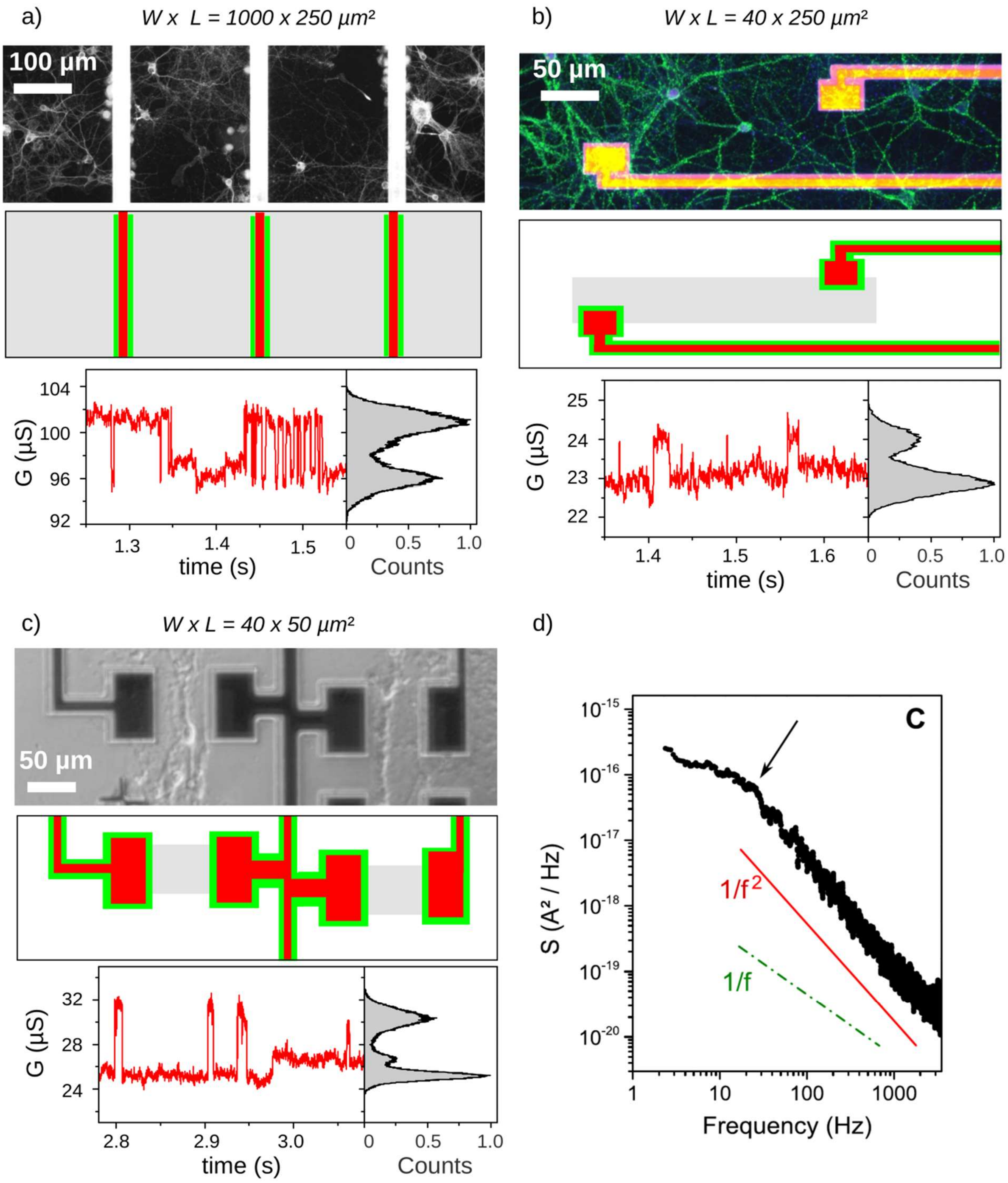

Figure 3.

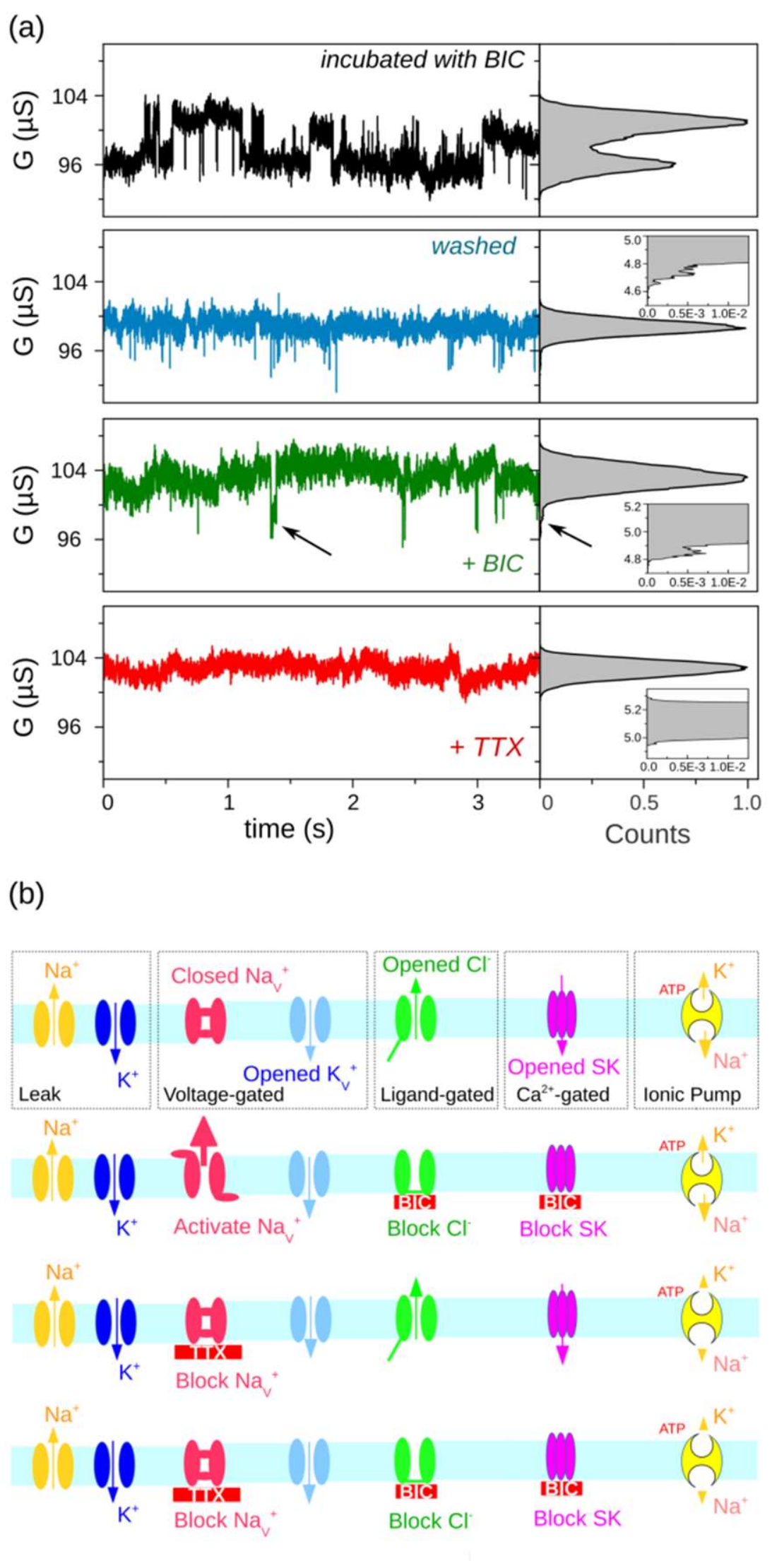

Figure 4.

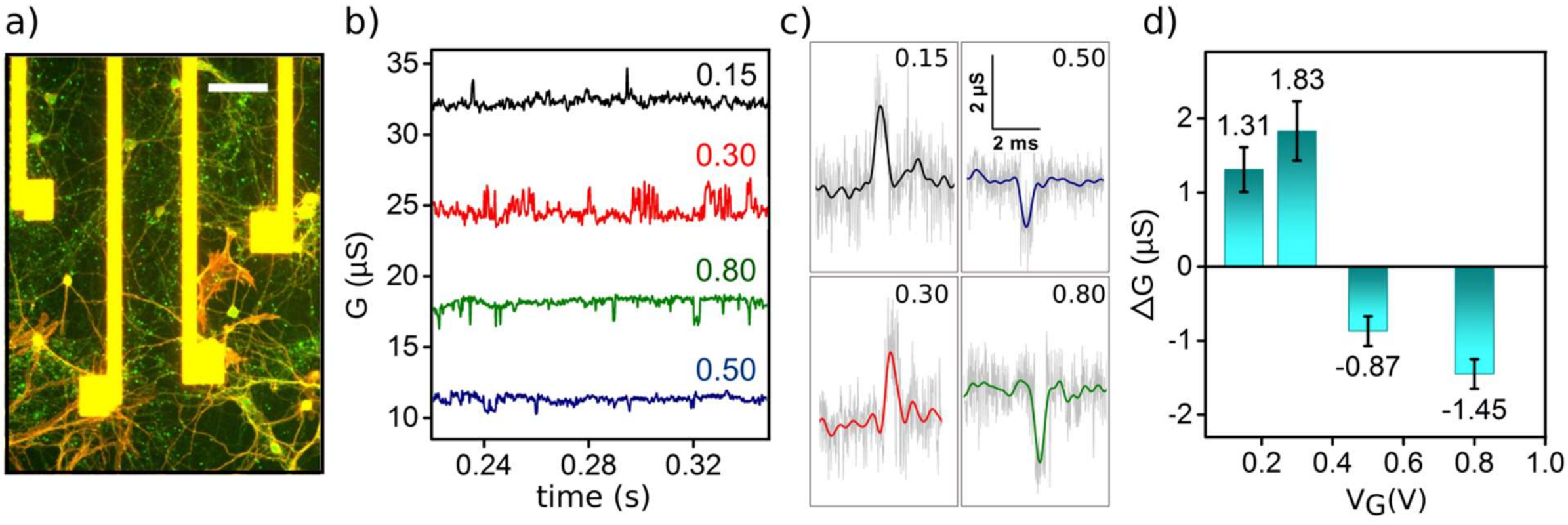

Figure 5.

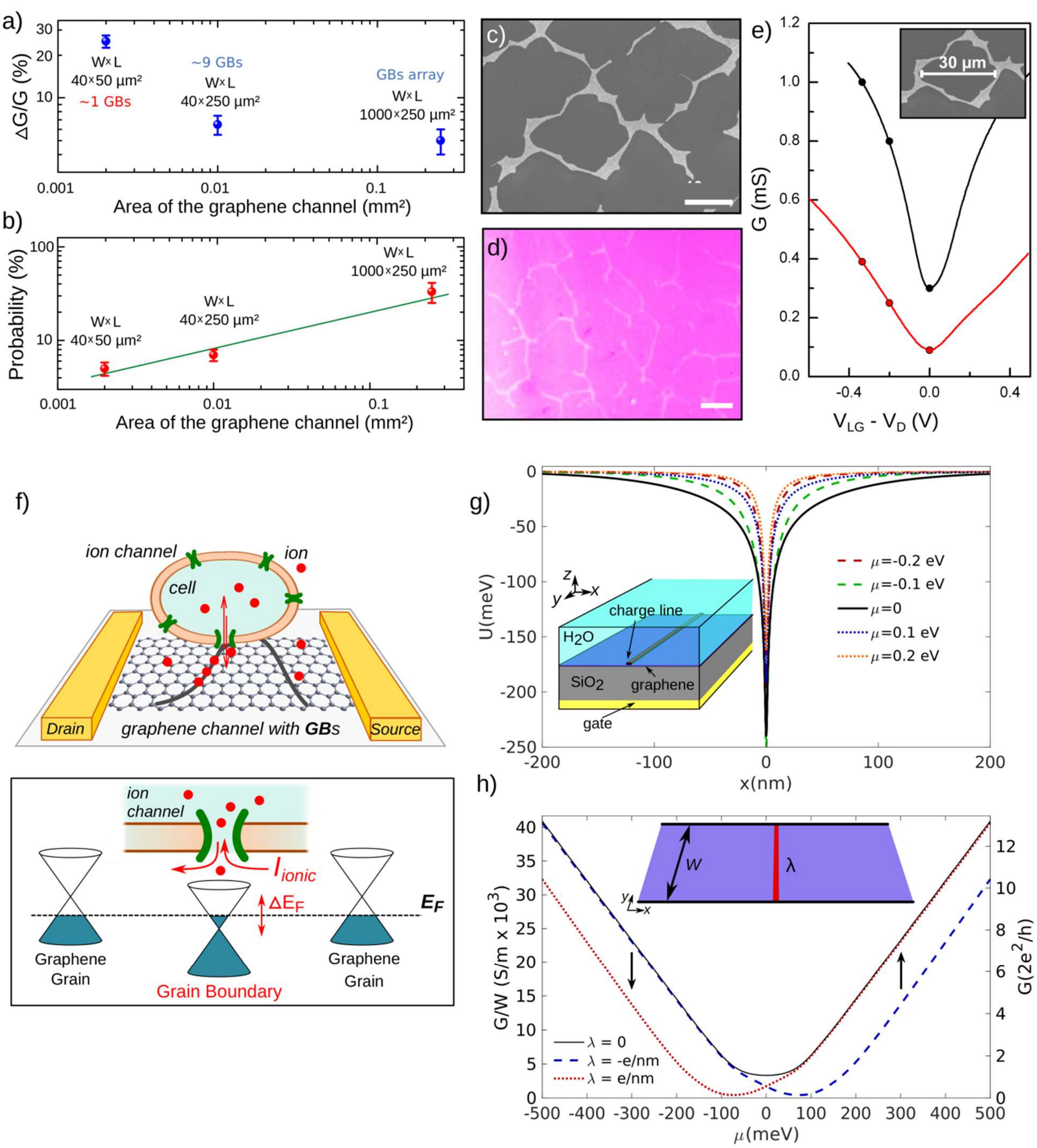

Figure 6.

ASSOCIATED CONTENT

**Supporting Information**. Supplementary figures S1 to S3

*Corresponding Author

E-mail: cecile.delacour@neel.cnrs.fr

ACKNOWLEDGMENTS

The authors thank G. Bres, J. L. Moncellin for electronics support. Z. Han, L. Marty and N. Bendiab for their support on the graphene growth. The authors gratefully acknowledge financial support from the University Joseph Fourrier (SMINGUE project), from la Région Rhône-Alpes (COOPERA project), and from the french Agence Nationale de la Recherche under the projects ANR-10-LABX-51-01 (Labex LANEF du Programme d'Investissements d'Avenir) the Lab Alliances on Nanosciences - Energies for the Future. ICN2 is supported by the CERCA programme/Generalitat de Catalunya. And the Spanish Ministry of Economy and Competitiveness and the European Regional Development Fund (Project No. FIS2015-67767-P MINECO/FEDER), MINECO (Grant No. SEV-2013-0295) and funded by the CERCA Programme/Generalitat de Catalunya. The authors acknowledge the Spanish Ministry of Economy and Competitiveness and the European Regional Development Fund (Project No. FIS2015-67767-P MINECO/FEDER),

ABBREVIATIONS

BIC Bicuculline, TTX Tetrodotoxin, PFA Paraformaldehyde.

REFERENCES

1. Ribrault, C., Sekimoto, K. and Triller, A. *Nature Reviews Neuroscience* **2011** 12(7), 375-387.

2. Ashcroft FM. *Nature* **2006**, 440:440–447.

3. Kim, J. B. *Korean journal of pediatrics* **2014** 57(1), 1-18.

4. Kostarelos, K., Novoselov, K. S. *Nat. Nanotechnol.* **2014,** 9, 744.

5. Hess, L. H., Jansen, M., Maybeck, V., Hauf, M. V., Seifert, M., Stutzmann, M., Sharp, D., Offenhäusser, A., Garrido, J. A. *Adv. Mater.* **2011,** 23, 5045-5049.

6. Cohen-Karni, T., Qing, Q., Li, Q., Fang, Y., Lieber, C. M. *Nano Lett.* **2010,** 10, 1098-1102.

7. Uesugi, E., Goto, H., Eguchi, R., Fujiwara, A., Kubozono, Y. *Sci. Rep.* **2013,** 3, 1595.

8. Schwierz, F. *Nat. Nanotechnol.* **2010,** 5, 487-496.

9. Koerbitzer, B., Krauss, P., Nick, C., Yadav, S., Schneider, J. J., and Thielemann, C. *2D Materials* **2016**, 3(2), 024004

10. Veliev, F., Han, Z., Kalita, D., Briançon-Marjollet, A., Bouchiat, V., and Delacour, C. *Frontiers in neuroscience* **2017**, 11, 466.

11. Lacour, S. P., Courtine, G., Guck, J. *Nature Reviews Materials* **2016** 1, 16063.

12. Veliev, F., Briançon-Marjollet, A., Bouchiat, V., Delacour, C. *Biomaterials* **2016,** 86, 33-41.

13. Fabbro, A., Scaini, D., Leon, V., Vázquez, E., Cellot, G., Privitera, G., Lombardi, L., Torrisi, F., Tomarchio, F., Bonaccorso, F., Bosi, S., Ferrari, A.C. *ACS Nano* **2015,** 10(1), 615-623.

14. Yang, K., Li, Y., Tan, X., Peng, R., Liu, Z. *Small* **2013,** 9 (9-10), 1492-1503.

15. Bendali, A., Hess, L. H., Seifert, M., Forster, V., Stephan, A. F., Garrido, J. A., Picaud, S. *Adv. Healthc. Mater.* **2013,** 2, 929-933.

16. Lorenzoni, M., Brandi, F., Dante, S., Giugni, A., Torre, B. *Sci. Rep.* **2013,** 3, 1954.

17. Park, S. Y., Park, J., Sim, S. H., Sung, M. G., Kim, K. S., Hong, B. H., Hong, S. *Adv. Mater.* **2010,** 23, 263.

18. Tang, M., Song, Q., Li, N., Jiang, Z., Huang, R., Cheng, G. *Biomaterials* **2013**, 34, 6402-6411.

19. Blaschke, B. M., et al. *2D Materials* **2017**, 4(2), 025040.

20. Hébert, C. et al. *Advanced Functional Materials* **2017,** 28(12), 1703976.

21. Kireev, D., Brambach, M., Seyock, S., Maybeck, V., Fu, W., Wolfrum, B., and Offenhäusser, A. *Scientific Reports* **2017**, 7(1), 6658.

22. Wang, Y. Y., Pham, T. D., Zand, K., Li, J., Burke, P. J. *ACS Nano* **2014,** 8(5), 4228-4238.

23. Graf, D., Molitor, F., Ensslin, K., Stampfer, C., Jungen, A., Hierold, C., Wirtz, L. *Nano Lett.* **2007,** 7, 238-242.

24. Han, Z., Kimouche, A., Kalita, D., Allain, A., Arjmandi-Tash, H., Reserbat-Plantey, A., Marty, L., Pairis, S., Reita, V., Bendiab, N., Coraux, J., Bouchiat, V. *Adv. Funct. Mater.* **2014,** 24, 964-970.

25. Fletcher, T. L., Cameron, P., De Camilli, P., Banker, G. *J. Neurosci.* **1991,** 11(6), 1617-1626.

26. Strassberg, A.F. and DeFelice, L.J. *Neural Comput.* **1993**, 5, 843–855

27. Isacsson, A., Cummings, A.W., Colombo, L., Colombo, L., Kinaret, J.M., Roche, S. *2D materials* **2017**, 4(1),1–13.

28. Huang, P. Y. et al. *Nature* **2011**, 469(7330), 389.

29. Seifert, M., Vargas, J. E., Bobinger, M., Sachsenhauser, M., Cummings, A. W., Roche, S., and Garrido, J. A. *2D Materials* **2015**, 2(2), 024008.

30. Salehi-Khojin, A., Estrada, D., Lin, K. Y., Bae, M-H., Xiong, F., Pop, E., Masel, R. I. *Adv. Mater.* **2012,** 24, 53-57.

31. Yasaei, P., Kumar, B., Hantehzadeh, R., Kayyalha, M., Beskin, A., Repnin, N.,Wang, C., Klie, R. F., Chen, Y. P., Kral, P., Salehi-Khojin, A. *Nat. Commun.* **2014,** 5, 4911.

32. Schedin, F., Geim, A. K., Morozov, S. V., Hill, E. W., Blake, P., Katsnelson, M. I., and Novoselov, K. S. *Nature materials* **2007**, 6(9), 652.

33. Sun, X., Wang, Z., and Fu, Y. Q. *Carbon* **2017**, 116, 415-421.

34. T. M. Radchenko, A. A. Shylau, I. V. Zozoulenko, and A. Ferreira *Phys. Rev. B* **2013**, 87, 195448.

35. Z. L. Mišković, and N. Upadhyaya *Nanoscale Res. Lett.* **2010**, 5, 505.

36. A. Cresti, R. Farchioni, G. Grosso, and G. Pastori Parravicini *Phys. Rev. B* **2003**, 68, 075306.

37. A. Cresti, N. Nemec, B. Biel, G. Niebler, F. Triozon, G. Cuniberti, and S. Roche *Nano Research* **2008**, 1, 361.

38. Kirton, M. J., Uren, M. J. *Adv. Phys.* **1989,** 38, 367–468.

39. Clément, N., Nishiguchi, K., Fujiwara, A., Vuillaume, D. *Nat. Commun.* **2010,** 1, 92.

40. Li, J., Pud, S., Petrychuk, M., Offenhausser, A., Vitusevich, S. *Nano Lett.* **2014,** 14(6), 3504-3509.

41. Sharf, T., Wang, N. P., Kevek, J. W., Brown, M. A., Wilson, H., Heinze, S., Minot, E. D. *Nano Lett.* **2014,** 14(9), 4925-4930.

42. Karnatak, P., Goswami, S., Kochat, V., Pal, A. N., and Ghosh, A. *Physical review letters* **2014**, 113(2), 026601.

43. Das Sarma, S., S. Adam, E. H. Hwang, and E. Rossi *Rev. Mod. Phys.* **2011**, 83, 407–470.

44. Kim, S., J. Nah, I. Jo, D. Shahrjerdi, L. Colombo, Z. Yao, E. Tutuc, and S. K. Banerjee *Applied Physics Letters* **2009**, 94(6), 062107.

## Supporting Information

# Sensing ion channel in neurons networks with graphene transistors

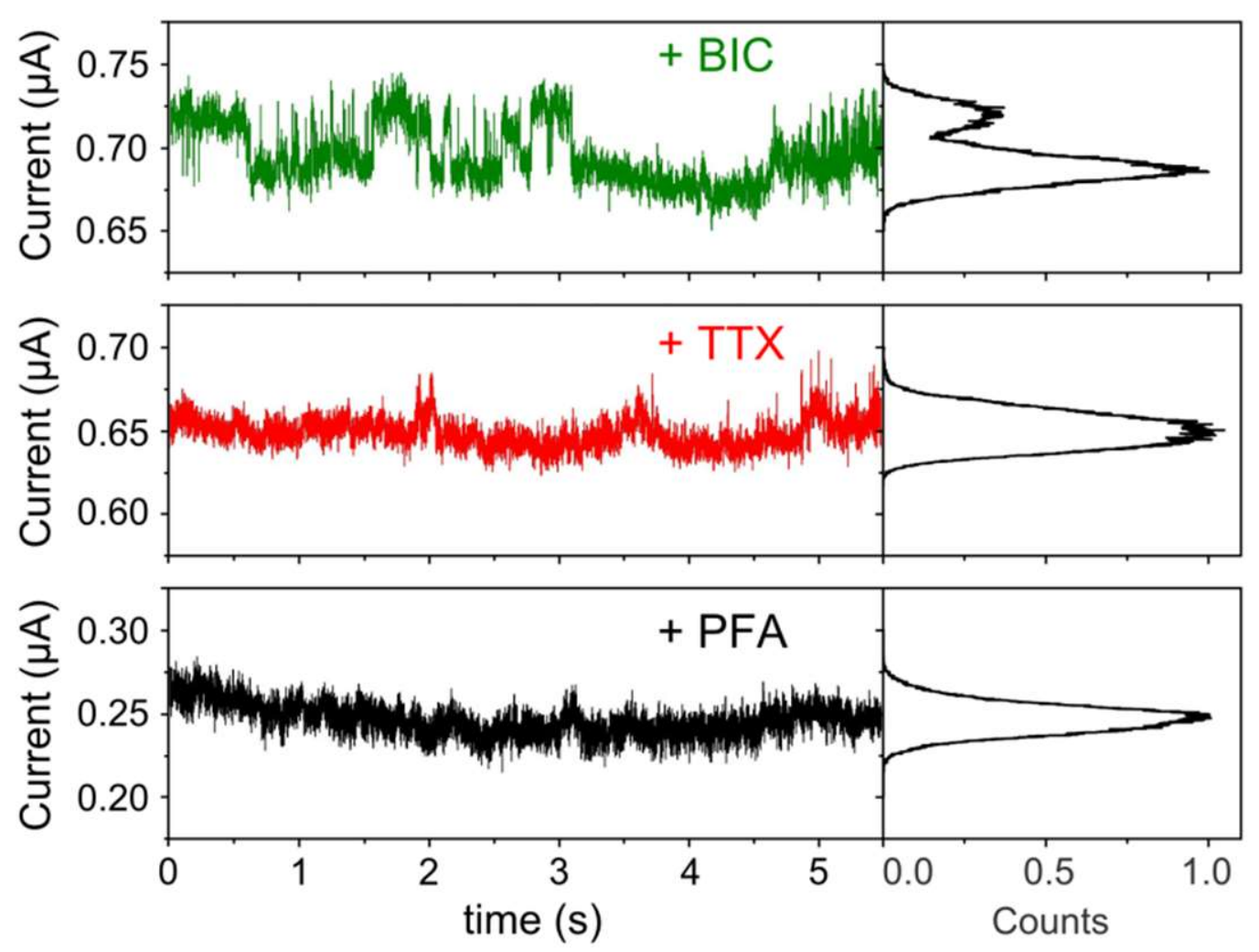

**Figure S1. Suppression of RTS by adding tetrodotoxin (TTX) to the extracellular medium.** Current traces (left) and corresponding histograms (right) obtained using a 40×250 μm² G-FET interfaced to cultured neurons (DIV19). The current traces are subsequently recorded at a constant bias voltage $V_{SD}$ = 50 mV and liquid gate potential $V_{LG}$ = 0.15 V. Clear two-state conductance fluctuations are observed when the cell culture medium is supplemented with bicuculline (BIC, top panel, green trace). The addition of tetrodotoxin (TTX, middle panel, red trace) completely suppresses the signals induced by neurons. Also paraformaldehyde, which is used to fix the neurons after the recordings, suppresses the signals (PFA, bottom panel, black trace).

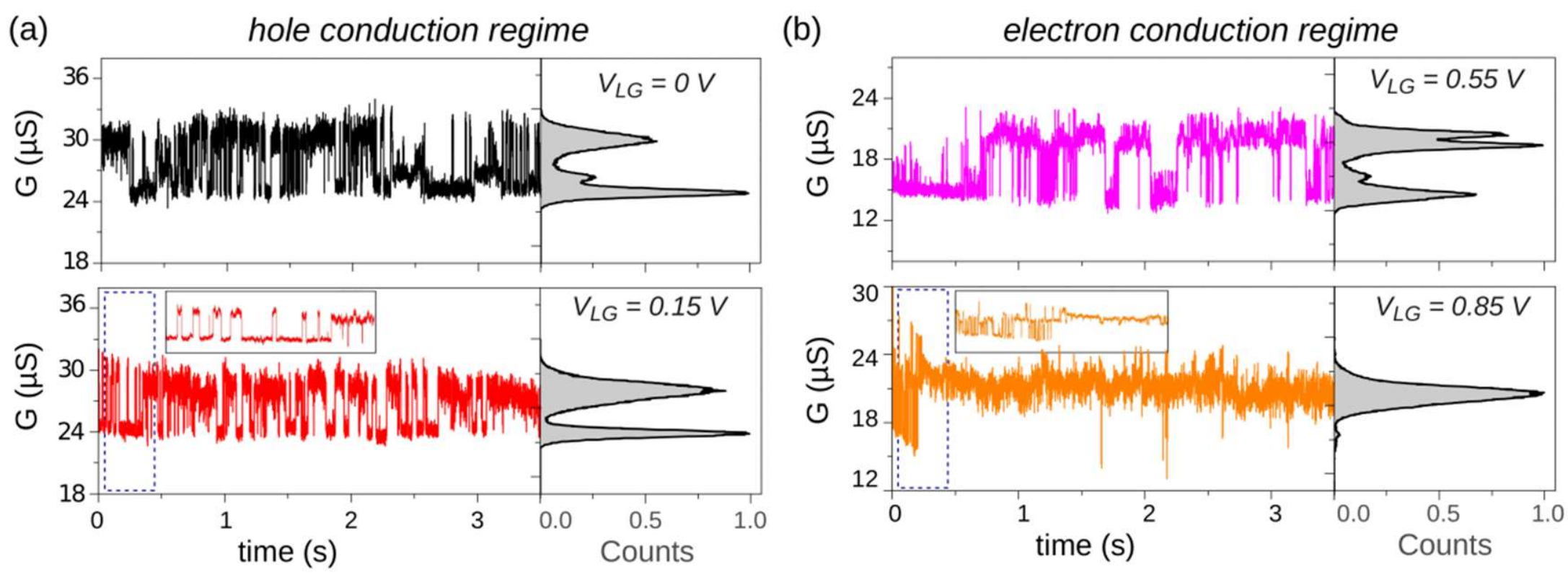

**Figure S2.** Conductance time traces and corresponding histograms recorded in the hole (a) and electron (b) operation regime of the G-FET (40×50 µm²). A zoomed view of the recorded traces is exemplarily demonstrated for the hole and electron operation regimes at $V_{LG}$ = 0.15V and 0.85V respectively. $V_{SD}$ = 30 mV during the drain-source current recording.

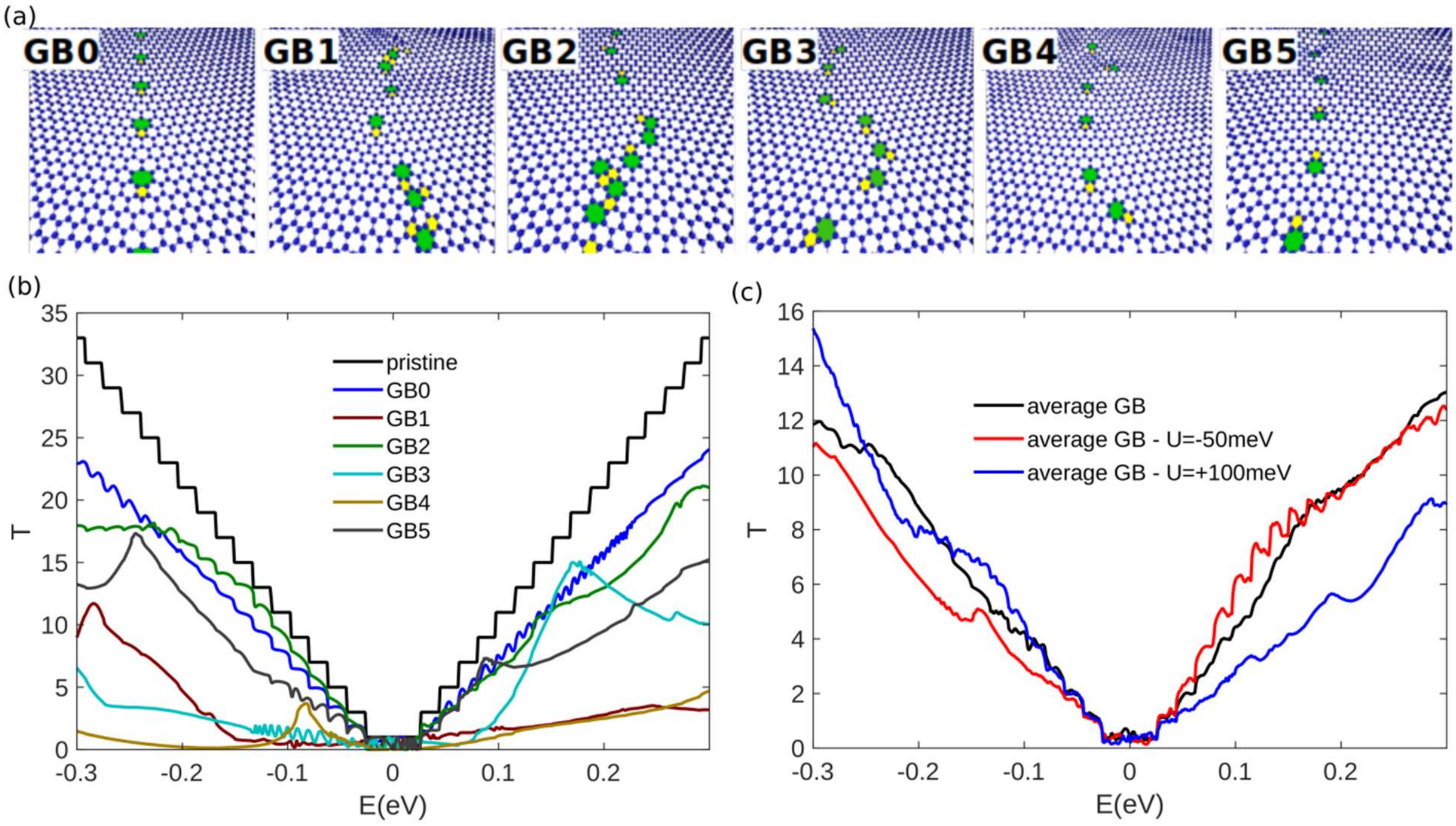

**FIGURE S3. Numerical simulations.** (a) Geometries of the considered grain boundaries. (b) Transmission coefficient for a 100 nm wide graphene ribbons with pristine geometry or in the presence of the different grain boundaries across its transverse section. (c) Transmission

coefficient averaged over the different GB geometries in the absence and in the presence of potential barriers with height $U$ over a 10 nm wide stripe along the GB.